\documentclass[conference]{IEEEtran}
\IEEEoverridecommandlockouts
\usepackage{cite}
\usepackage{amsmath,amssymb,amsfonts}
\usepackage{algorithmic}
\usepackage{graphicx}
\usepackage{textcomp}
\usepackage{xcolor}
\usepackage[nolist,nohyperlinks]{acronym}
\usepackage{hyperref}
\usepackage{booktabs}
\usepackage{lipsum}
\usepackage{footnote}
\usepackage{mathrsfs}
\makesavenoteenv{tabular}
\usepackage{colortbl} 
\definecolor{lightgray}{gray}{0.9} 
\def\BibTeX{{\rm B\kern-.05em{\sc i\kern-.025em b}\kern-.08em
    T\kern-.1667em\lower.7ex\hbox{E}\kern-.125emX}}
\begin{document}

\title{SHIFT: An Interdisciplinary Framework for Scaffolding Human Attention and Understanding in Explanatory Tasks
\thanks{$^{1}$Medical Assistance Systems, Medical School OWL and $^{2}$Center for Cognitive Interaction Technology, CITEC from Bielefeld University, Germany. $^{*}$Corresponding Author: André Groß \tt\small agross@techfak.uni-bielefeld.de.}%
\thanks{This research was funded by the Deutsche Forschungsgemeinschaft (DFG, German Research Foundation): TRR 318/1 2021-438445824 “Constructing Explainability”.}
}

\author{\IEEEauthorblockN{André Groß$^{1,2*}$, Birte Richter$^{1,2}$ and Britta Wrede$^{1,2}$}
}

\begin{acronym}[list of abbreviations]
\acro{hri}[HRI]{\textit{Human-Robot Interaction}}
\acro{xai}[XAI]{\textit{Explainable Artificial Intelligence}}
\acro{ros}[ROS]{\textit{Robot-Operating-System}}
\acro{xml}[XML]{\textit{eXtensible-Markup-Language}}
\acro{scxml}[SCXML]{\textit{State Chart eXtensible-Markup-Language}}
\acro{tcp}[TCP]{\textit{Transmission Control Protocol}}
\acro{its}[ITS]{\textit{Intelligent Tutoring Systems}}
\acro{mdp}[MDP]{\textit{Markov Decision Process}}
\acro{pomdp}[POMDP]{\textit{Partial Observable Markov Decision Process}}
\acro{gui}[GUI]{\textit{Graphical User-Interface}}
\acro{llm}[LLM]{\textit{Large Language Model}}
\acro{fsm}[FSM]{\textit{Finite-State-Machine}}
\acro{rl}[RL]{\textit{Reinforcement-Learning}}
\acro{bn}[BN]{\textit{Bayesian Networks}}
\acro{tva}[TVA]{\textit{Theory of Visual Attention}}
\acro{crc}[CRC]{\textit{Collaborative Research Centre}}
\acro{ai}[AI]{\textit{Artificial Intelligence}}
\acro{aoi}[AOI]{\textit{Area of Interest}}
\end{acronym}

\maketitle

\begin{abstract}
In this work, we present a domain-independent approach for adaptive scaffolding in robotic explanation generation to guide tasks in human-robot interaction.
We present a method for incorporating interdisciplinary research results into a computational model as a pre-configured scoring system implemented in a framework called SHIFT.
This involves outlining a procedure for integrating concepts from disciplines outside traditional computer science into a robotics computational framework.
Our approach allows us to model the human cognitive state into six observable states within the human partner model.
To study the pre-configuration of the system, we implement a reinforcement learning approach on top of our model.
This approach allows adaptation to individuals who deviate from the configuration of the scoring system.
Therefore, in our proof-of-concept evaluation, the model's adaptability on four different user types shows that the models' adaptation performs better, i.e., recouped faster after exploration and has a higher accumulated reward with our pre-configured scoring system than without it.
We discuss further strategies of speeding up the learning phase to enable a realistic adaptation behavior to real users.
The system is accessible through docker and supports querying via ROS.
\end{abstract}


\section{Introduction}

\begin{figure}[!t]
    \centering
    \includegraphics[width=\linewidth]{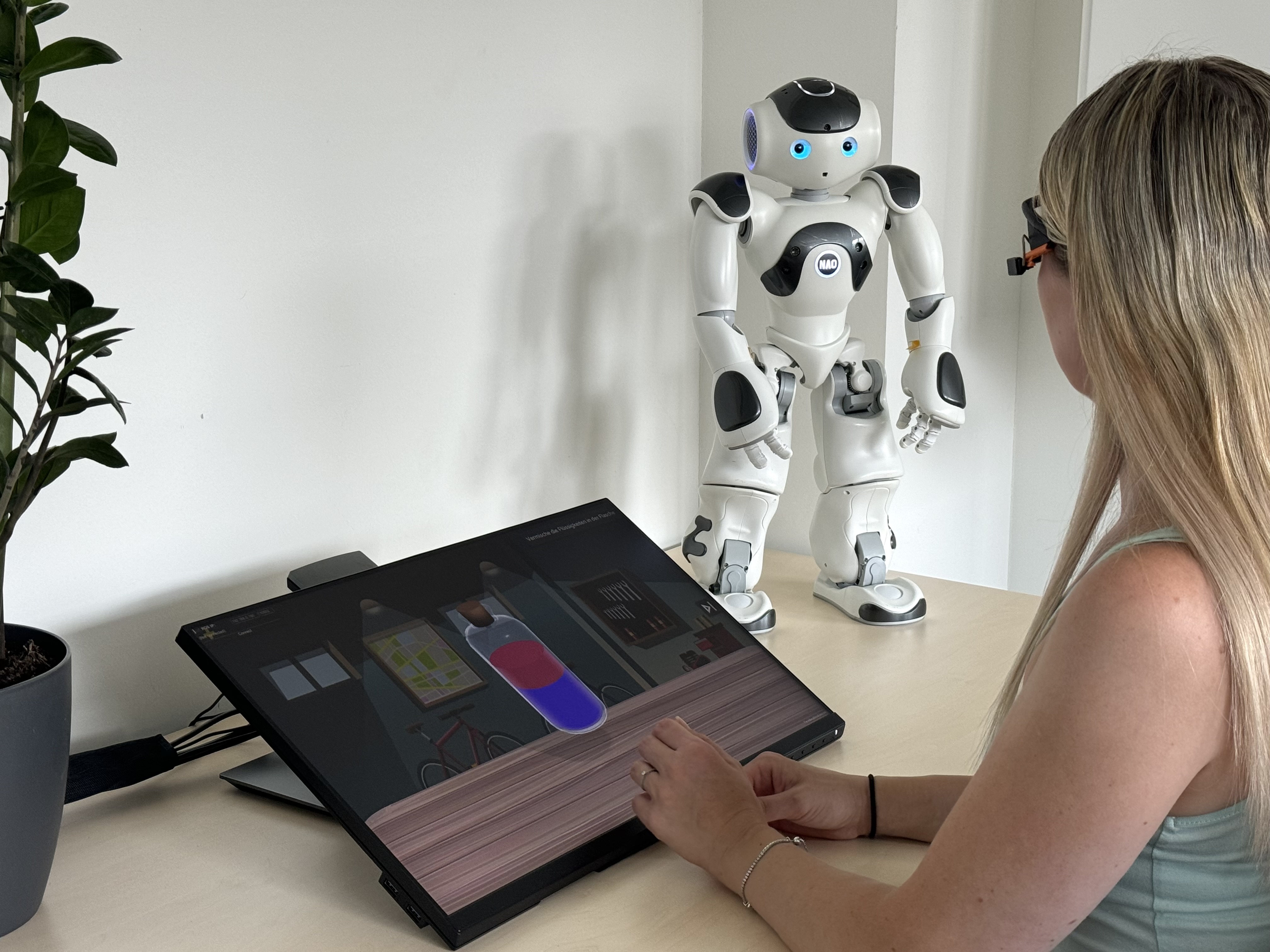}
    \caption{Interaction between human and robot in a social context. The robot using individual scaffolding strategies to guide during task solving.}
    \label{fig:human_robot_interaction_intro}
    \vspace{-18pt}
\end{figure}

Recent advancements in \ac{ai} and robotics have made the development of robots to assist with everyday manipulation tasks increasingly achievable. 
A key area of research focuses on enabling human-like interaction with objects across a wide range of tasks, from food preparation to object assembly \cite{sheridan2016human}.
For this purpose, cognitive architectures have been proposed that enable robots to plan sequences of actions for specific tasks \cite{beetz2010cram}.
This enabled the community to create various components for these architectures, addressing not only object manipulation but also the interaction of robots with humans.
Various \ac{hri} studies (e.g., \cite{wang2018facilitating, wang2018human}) are investigating, how a robot and a human can solve tasks together.
Collaborative task solving is supported not only by physical interaction, but also by multimodal verbal communication \cite{tatarian2022does}.
While dialogue managers in robotics \cite{Carlmeyer2014, ren2015tfsm} are useful for managing the flow of conversations by focusing on turn-taking and content delivery, they fall short when it comes to generating nuanced, context-sensitive explanations that speak to the cognitive level of the human interlocutor in domain-dependent interactions \cite{zhao2019review}.
\acp{llm} on the other hand, also reproduce such verbal interactions, but they are not transparent \cite{KASNECI2023102274}.
The process of reconstructing social practices \cite{rohlfing2020explanation}, like the reconstruction of human behavior, is still relatively unexplored.
Fundamental \ac{hri} research is essential to understand the cognitive processes in task-related verbal interactions, enabling robots to generate explanations and communicate effectively with humans in everyday life.
Therefore, the goal of this work is to propose an approach for \textbf{adaptive scaffolding when explaining a task} to a human (\autoref{fig:human_robot_interaction_intro}).
We bring together interdisciplinary insights from (I) linguistics on verbal \textbf{scaffolding strategies}, (II) psychology on \textbf{monitoring} and modeling human attention, and (III) robotics on task awareness as a metric of understanding, deriving a nuanced \textbf{computational adaptation model} for explaining tasks.
In our approach, we formalized linguistic and psychological insights to generate a \textbf{configuration} for a \textbf{scoring system} that determines the appropriate scaffolding strategy for specific conditions.
As we don't know if the resulting configuration is correct, we implemented a \ac{rl} algorithm that allows to research on an optimal configuration of the system.
Through the \ac{rl} approach, adaptation to individual deviations from the norm is possible, and we investigate their impact on learning efficiency.

\section{Related Work}

\subsection{Decision-Making in Social Robotics Tutoring}

Social robotics focuses on facilitating interactions between humans and robots at a social level \cite{kanda2017human}, while \ac{its} bridges the gap between robotics and social interaction \cite{almasri2019intelligent}, aiming to assist humans in task-solving with robotic (or \ac{ai}-) guidance.
The challenge in modeling social interactions lies in designing representations of real-world settings for the robot.
To achieve this, \ac{its} distinguishes between models of the student, the domain and pedagogical models \cite{almasri2019intelligent}.
The student model monitors human behavior and represents the current knowledge of the student.
The pedagogical model selects appropriate actions and adapts the planning process based on the student model.
The domain model outlines fundamental rules about the environment and establishes the overall guidelines for interactions between the explainer and the explainee.
In~\cite{ramachandran2019personalized} an \ac{its} was developed, featuring the NAO robot \cite{naogouaillier2009mechatronic}, which was designed to assist with solving mathematical problems.
The main challenge is to address the decision-making process using human feedback, such as engagement, to determine the most effective actions for supporting the human in solving the math problem.
In \ac{hri}, it is often difficult to design states, state transitions, or decision problems in general that accurately reflect the problem.
Various approaches \cite{martins2019alphapomdp, RAMIREZNORIEGA20171488}  for \ac{hri} and action selection have been proposed to address the lack of data in the interaction with humans using computational models like \acp{fsm} \cite{carlmeyer2018hesitating}, \acp{mdp} \cite{folsom2013tractable} or \ac{bn} \cite{kopp2018conversational} to formalize these types of problems.
But human behavior and the human's current understanding of the world are difficult to read.
Finding rules for real-world problems, such as how to explain something to someone in the best possible way, is intuitive for humans, but unsolvable by default for robots that have no prior knowledge of the domain nor the human.
The problem is particularly difficult to formalize when the state transitions and probabilities for choosing actions in the model for the domain are not yet known \cite{anwar2022applying}.
In such non-stationary problems, a transition function cannot necessarily be derived, and other methods have to be used to solve the decision-making problem.
In social robotics, robots need to learn from the interactions they have with humans.
Based on user feedback received during these interactions, \acl{rl} is used to adapt \cite{wang2018reinforcement}.
In real-world settings, the number of interactions with humans is also highly limited.
To reduce the state space with this knowledge \cite{barraquand2008learning}, models can be preconfigured with probabilities from previous studies or expert knowledge.
However, this prior knowledge does not exist for environments that have not yet been researched.

\subsection{Adaptive Scaffolding in Explanation Generation}

The \textit{Transregional} \ac{crc} 318 “Constructing Explainability” \cite{trr318_overall} investigates what explanations are needed in \acl{hri} and how to generate such explanations in an adaptive way.
A basic structure for the design of explainable architectures has been proposed in \cite{booshehri2024towards}.
\cite{icaart23} have implemented a system for the generation of adaptive explanations in game domains. 
Most research focuses on adapting such task-depending strategies.
However, there is little research on how scaffolding strategies are applied independently across tasks.
Linguistic strategies are particularly suitable here.
In \cite{gros_scaffolding_nodate}, negations have been shown to be an effective scaffolding strategy, yielding better action performance than affirmative explanations.
In~\cite{richter2023eeg, richter2021attention} hesitations have been used as a scaffolding strategy in a human-robot tutoring scenario and successfully regained the human's attention yielding better retention.
These explanation strategies differ only on the linguistic level.
The distinction between WHAT and HOW a robot should explain is one important focus of research \cite{klein2009erklaren}.
Even if the scaffolding strategies are task-independent, the rules for when a strategy should be applied are defined by a scoring system.
In some cases, they are defined at predefined points in the interaction (e.g., \cite{gros_scaffolding_nodate, richter2023eeg}), or depending on human behavior (e.g., \cite{richter2021attention}).
However, as far as we know, no approach currently exists to adapt scaffolding strategies to the user.

\subsection{Research Scope}

Our goal is to introduce a system that guides humans through explanatory dialogues and provides scaffolding actions to support task-solving based on individual needs.
Therefore, we present a generic, domain-independent computational model that departs from traditional decision problem formalizations.
Our model adaptively selects scaffolding strategies through a scoring system approach that incorporates a novel operationalization of human attention.
A web-based \ac{gui} allows monitoring, and the model is allowed to learn (and thereby adapt its configuration of the scoring system) from human feedback through \ac{rl}.
This setup is supposed to improve human task performance while maximizing the robot's rewards during interactions.
It integrates components for monitoring human behavior, decision-making, and robotics, along with a service that offers customized scaffolding strategies based on environmental observations.

\section{Computational Model for Adaptive
Scaffolding when Explaining a Task}

In the context of explanatory dialogues between humans and robots, the targeted use of explanatory strategies lead to better and more understandable interaction.
For this purpose, explanatory strategies should be based on the individual's understanding of and attention to the task.
To achieve this, the robot must be able to recognize the human's cognitive state, i.e., perception of the task.
We suggest the following states to be indicative of the human's cognitive state:
(1) the general processing capacity; (2) the perception of stimuli from the environment as the distribution of the visual focus of attention; (3) the history of previous interactions.
It must be emphasized here that we do not try to operationalize this cognitive state for a certain task or context, but explicitly map the cognitive state of the human in the context of “solving tasks”.
We model this problem as a classical decision-making model in \ac{mdp} form.
A \ac{mdp} is defined as a 4-tuple consisting of states, actions, a probability transition function and a reward model \cite{puterman1990markov}.

\subsection{State-Space: Observations for Monitoring}

We operationalize the model's state-space through the combination of attentional parameters and the history of interactions in tasks.
From the \ac{tva} \cite{bundesen1990theory}, we derive the visual capacity ($C$) and the distribution of attention between targets by their weights ($W$).
To apply these theoretical concepts in real-world settings, we develop a configuration for the scoring system around the attention parameter for our measurements.
By integrating the user's attention with the task awareness, we generate observation triples to serve as our states.

\subsubsection{Processing Capacity} \label{sec:processing_capacity}

The processing capacity ($\mathcal{C}$), inspired by the \ac{tva} capacity \cite{scharlau2020effects}, represents a kind of battery that the human partner possesses during the interaction.
When the human receives a certain type of verbal scaffolding strategy, their capacity decreases based on the cognitive effort required to process it.
By receiving repeated verbal strategies as actions ($\mathcal{A}' = \mathcal{A}$), the battery recovers over time by $\Delta_{rep}$.
This observation type is grouped by absolute thresholds into the classes $\mathcal{C} = \{\textit{low}, \textit{high}\}$.
It describes if there is enough capacity to process a cognitively demanding ($\Delta_{cog}$) or less cognitive demanding ($\Delta_{ncog}$) scaffolding strategy next. $\mathcal{C}'$ describes the new processing capacity after processing.
\vspace{-10pt} 

\begin{displaymath}
\mathcal{C}' = 
\begin{cases} 
\min(\mathcal{C} + \Delta_{rep}, \mathcal{C}_{max}), & \text{if } \mathcal{A}' = \mathcal{A} \\
\max(\mathcal{C} - \Delta_{cog}, \mathcal{C}_{min}), & \text{if } \mathcal{A}' \text{ is cog. demanding} \\
\max(\mathcal{C} - \Delta_{ncog}, \mathcal{C}_{min}), & \text{otherwise}
\end{cases}
\label{eq:capacity}
\end{displaymath}

\subsubsection{Visual Distribution of Attention}

The gaze distribution ($w$) is derived from the \ac{tva}'s weights \cite{Banh_Tünnermann_Rohlfing_Scharlau} represents the distribution of gaze behavior across individual target objects in the environment ($\mathcal{C}_{max} = \sum_{i} w_i$).
These targets are embedded within the tasks and outline \ac{aoi}, where highlighting these helps in task completion.
By fixating objects via gaze ($i = i_f$), the distribution of visual attention shifts to the target stimulus ($\Delta_w$).
The weights for unfixated targets ($i \neq i_f$) is partitioned by an excess function.
The distributions are decaying (normalized by $\Delta_{w_{new}}$). $w'$ describes the new distribution after processing.
\begin{displaymath}
\begin{array}{ll}
w_i' = \begin{cases} 
\min(w_{i} + \Delta_w, \mathcal{C}_{max}), & \text{if } i = i_f \\
\max\left(w_{i} - \frac{\Delta_{w_{new}}}{N - 1}, w_{min} \right), & \text{if } i \neq i_f
\end{cases} \\
\\
\text{where } \Delta_{w_{new}} = \min(w_{i_{f}} + \Delta_w, \mathcal{C}_{max}) - w_{i_f}
\end{array}
\label{eq:weights}
\end{displaymath}

The frequency of focus shifts ($\mathcal{FS}$) and the last focused object ($\mathcal{F}_{last}$) are considered in addition to the distribution of gaze behavior.
The focus shift increases by $\Delta_{inc}$ when the last focused object is not fixed ($\mathcal{F}_{last} \neq i_f$).
It decreases by $\Delta_{dec}$ when the focus is kept ($\mathcal{F}_{last} = i_f$).
\begin{displaymath}
\mathcal{FS}' = 
\begin{cases}
\min(\mathcal{FS} + \Delta_{inc}, \mathcal{FS}_{max}), & \text{if } \mathcal{F}_{last} \neq i_f \\
\max(\mathcal{FS} - \Delta_{dec}, \mathcal{FS}_{min}), & \text{if } \mathcal{F}_{last} = i_f
\end{cases}
\label{eq:focus_shift}
\end{displaymath}

By considering the distribution of the viewing behavior and the frequency of the change of focus, the visual attention of the user is classified in the groups $\mathcal{W} = \{\textit{distracted}, \textit{uncertain}, \textit{focused}\}$ by fixed thresholds.

\subsubsection{Task Awareness}

The history of interactions is highly relevant during the use of scaffolding strategies.
Especially when an expectation is already anchored in the interaction partner's intention and the robot wants to contradict it, the use of contrastive explanatory strategies is highly positive \cite{gros_scaffolding_nodate}.
Accordingly, we record the history of performances from the human and store how these tasks were solved.
To better address the question “how is a task solved”, we divide the task understanding into two dimensions.
The dimension of the understanding of a task at the theoretical level (comprehension), and the practical implementation of the solution to the task (enabledness) \cite{buschmeier2023formsunderstandingxaiexplanations}.
This differentiation allows for the use of scaffolding strategies that are more focused and supportive of practical action implementation.
By taking the task performance into account, we group the task awareness for each task in the classes $\mathcal{T} = \{\textit{unknown}, \textit{ failure}, \textit{ misc. enabl.}, \textit{ misc. compreh.}, \textit{ success}\}$.
The grouping takes place via the ratio of the task performance on the two dimensions.
In the group of $\text{failure}$, the tasks are considered incorrect on both dimensions.
By a double positive task-performance, the task is $\text{success}$ and in different results a $\text{misconception}$.
Considering our observable parameters, the state-space ($\left| \mathcal{S}_{monitoring} \right| = 30$) is represented as triples.
\vspace{-12pt}

\begin{displaymath}
\mathcal{S}_{monitoring} = \{(c, w, t) \mid c \in \mathcal{C}, w \in \mathcal{W}, t \in \mathcal{T}\}
\label{eq:states}
\end{displaymath}

\subsection{Actions: Explanation Scaffolding Strategies}

A set of actions needs to be selected along with the selection of explanatory strategies based on the observation triples.
The system described in this article allows configuring these actions in a flexible way.
For the model's introduction, we initially focus on negation and hesitation.
Linguistic negation is a grammatical mechanism used to deny or reject a statement. It is frequently employed in everyday situations to counter positive assertions \cite{beltran2021inhibitory}, thereby creating a contrast with a preceding concept.
For instance, instructional signs often clarify actions by correcting common mistakes, such as glass doors being pushed instead of pulled, with messages like "Pull, don’t push!" \cite{singh2023contrastiveness, gros_scaffolding_nodate}.
Negation strategies are implemented in our model with the types $\mathcal{N} = \{\textit{affirmation}, \textit{negation+affirmation}, \textit{negation}\}$.
In our configuration of the scoring system, they are used to correcting misconceptions or errors in the execution of a task, thereby addressing task awareness and the visual distribution of attention.
Hesitations, on the other side, give the listener time to process the information already provided or to prepare a more detailed response by pausing \cite{collard2009disfluency} and could be useful to regain the attention of a distracted user and improve the task performance~\cite{richter2021attention}.
Hesitation actions are described by $\mathcal{H} = \{\textit{none}, \textit{hesitation}\}$.
They are intended to address processing capacity issues.
Both strategies have various options in their form of explanation.
A combination of those strategies has not been thoroughly researched yet.
Negations are our primary scaffolding strategy and always appear in the explanation process.
All other modalities, such as hesitations in this example, are in addition to negation.
Thus, we focus on the following actions ($\left| \mathcal{A} \right| = 6$) in our model for the action selection process.
\vspace{-14pt}

$$
\mathcal{A} = \{(\textit{negation}, \textit{hesitation}) \mid \textit{negation} \in \mathcal{N}, \textit{hesitation} \in \mathcal{H}\} \\
$$

\subsection{State Reduction: Cognitive States of the Partner Model}

The use of decision problems in human interaction presents a challenge in social robotics.
Simulating an unlimited number of interactions is impractical, requiring a limited number of iterations for decision-making algorithms.
To address this, we reduce the state space through a pre-configured scoring system in a pre-processing step.
The literature of social robotics does not yet contain much on the adaptive use of explanatory strategies in verbal interaction between humans and robots.
This is particularly challenging because we lack direct prior knowledge of how different parameter combinations will predictably influence scaffolding strategy outcomes.
However, we abstract recommendations from previous studies and literature about which explanatory strategy to prefer for each observation value standalone.
We incorporate an externally configurable scoring system ($\mathcal{H}_{\mathcal{S}}(o)$) into our model (\autoref{tab:scoring_list}).
\vspace{-6pt} 
\begin{table}[!h]
\renewcommand{\arraystretch}{1.3}
\caption{Scoring System for Scaffolding Strategy Generation ($\mathcal{H}_\mathcal{S}$).}
\centering
\begin{tabular}{ll|ccc}
\toprule
Monitoring & Observation & Negation & Hesitation & ...\\
\hline
Processing Capacity & Low & 0 & 1 & .. \\
Processing Capacity & High & 1 & 0 & .. \\
Gaze Distribution & Distracted & 1 & 1 & .. \\
Gaze Distribution & Uncertain & 1 & 1 & .. \\
Gaze Distribution & Focused & 0 & 0 & .. \\
Task Awareness & Unknown & 0 & 1 & .. \\
Task Awareness & Failure & 0 & 1 & .. \\
Task Awareness & Misc. Enabl. & 1 & 0 & ..\\
Task Awareness & Misc. Compr. & 0 & 1 & ..\\
Task Awareness & Success & 1 & 0 & ..\\
.. & .. & .. & .. & ..\\
\bottomrule
\end{tabular}
\vspace{-6pt} 
\label{tab:scoring_list}
\end{table}

The goal of the scoring system is the calculation of an explanation score ($\textit{Scaffolding}_{score}$) based on the classified observations ($o \in \textit{Observations}$) for each strategy.
The score describes the need for the respective explanatory strategy considering the corresponding triplet.
For a triple, you get one score of each explanatory strategy.
The model also allows that these scores can be weighted by a scaffolding factor ($S_{weight}$) during runtime.
For example, the observation triple $(c, w, t)$ leads to the negation score A and the hesitation score B.
These scores indicate a percentage of how much of the explanatory strategy should be used for this particular situation of observations.
\vspace{-3pt} 
\begin{displaymath}
\textit{Scaffolding}_{score} = S_{weight} \cdot \left( \frac{\sum_{o \in \textit{Observations}} \mathcal{H}_{\mathcal{S}}(o)}{S_{max}} \right)
\label{eq:scaffolding_score}
\end{displaymath}

These explanation scores are then grouped by a threshold-based ranking and describe the need for a scaffolding strategy in the respective set of observations.
However, by calculating the explanation scores for each scaffolding strategy and the concatenation of strategies from our action space, we establish a ground truth based on our scoring system, specifying which action should be selected in each state.
This scoring system enables us to reduce the previous 30 states to 6 new states, represented by cognitive states of the human.
\vspace{-1pt} 
\begin{align}
\mathcal{S}_{cog} = \{\notag & \textit{Engaged Observer} \textit{ (affirmation)}, \\ 
      \notag & \textit{Engaged Misinterpreter} \textit{ (negation+affirmation)}, \\ 
      \notag & \textit{Distracted Misinterpreter} \textit{ (negation)}, \\ 
      \notag & \textit{Overwhelmed Struggler} \textit{ (affirmation \& hesitation)}, \\ 
      \notag & \textit{Unfocused} \textit{ (negation+affirmation \& hesitation)}, \\ 
      \notag & \textit{Uncertain} \textit{ (negation \& hesitation)}\}
\end{align}

Other models rely on varying assumptions for different user types \cite{icaart23}, while in our approach, the newly defined state space encodes the different human states in the same user.
The \textit{engaged observer} is a state where the human has a slightly increased capacity but is always focused by gaze behavior.
\textit{Engaged misinterpreter} describes high capacity, a misconception on enabledness dimension, or the full success in solving the task in the past.
\textit{Distracted misinterpreter} is similar to the \textit{engaged misinterpreter} but is distracted or uncertain in gaze behavior.
\textit{Overwhelmed struggler} represents the human with low capacity and an unknown or wrong task awareness of the task.
It describes a misconception on the comprehension side.
\textit{Unfocused} describes humans with distracted or uncertain gaze focus with a slight increase of processing capacity.
The state \textit{Uncertain} describes a full need of negation and hesitation, which is not in the initial configuration of the model, but it could become real by adjusting the action factors during runtime.

\subsection{Decision-Making: Adaptive Scaffolding}

\begin{figure*}[!t]
      \centering
      \includegraphics[width=\textwidth]{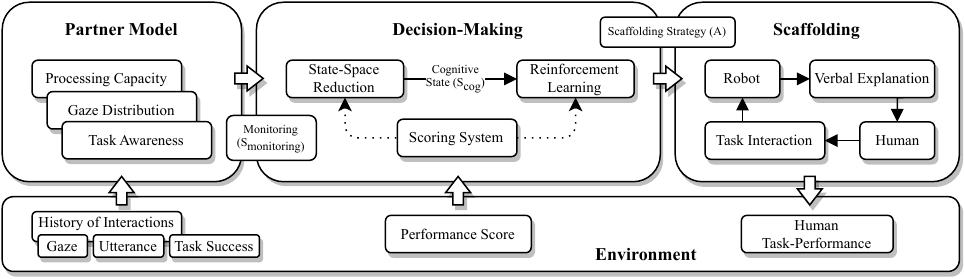}
      \vspace{-18pt} 
      \caption{The computational model of SHIFT with information flow. From the monitoring of the partner model of the human through the decision-making system to the generation of scaffolding actions for the \acl{hri}.}
      \label{fig:model_overview}
      \vspace{-16pt} 
\end{figure*}

In real-world tasks, the environment dynamics change over time.
Making decisions in such environments is challenging for decision-making problems because the environment does not adhere to a consistent set of rules.
In our approach, we aim to model human cognitive states in a more general way, yet the challenges persist.
Human behavior is difficult to predict, and the transition function of our states remains completely unexplored.
In such a non-stationary \cite{akalin2021reinforcement} environment, \ac{rl} agents interact with their environment and receive feedback directly during the interaction.
Actions which yield to negative feedback for the agent could produce a penalty, and positive feedback can be rewarded positively.
With our highly configurable approach, we aim to implement a method for evaluating our configuration of the scoring system during interaction by applying a reinforcement learning (\ac{rl}) algorithm on top of our model.
In this context, we are integrating a model-free learning algorithm into our system.
We aim to identify optimal scaffolding strategies for the human's cognitive states.
Therefore, we implement a Q-learning algorithm, with the use of the bellman-equation \cite{watkins1992q} on top of our model.
The Q-learning approach is used initially as an example of \ac{rl} and can be substituted with other methods.
We use our previous calculations to initialize a Q-table with pre-existing action suggestions, avoiding starting with zero Q-values.
Our scoring system, informed by literature and expert knowledge, serves as prior knowledge for the learning agent.
With this prior knowledge and a reduced state space, keeping our Q-table small, we make decisions while interacting with people in just a few iterations.
The learning algorithms select an action based on a decayed-epsilon-greedy algorithm \cite{tokic2010adaptive}.
At the beginning of learning, the agent explores more often.
Over time, the exploration decreases and the exploitation phases begins, and the agent prefers the actions with the highest Q-value from the table — the exploration decays.
Additionally, to ensure an effective exploration phase, the learning algorithm prioritizes selecting all unvisited actions first rather than choosing exploration actions randomly \cite{castro2014learning}.

\subsection{Reward: User's Task-Performance}

Feedback from the environment is essential for our model to learn and adapt to user behaviors.
Since our system is designed to assist humans in task-solving across different domains, a standardized interface is necessary as a reward function.
From the guidance of the robot through the use of actions (scaffolding strategies), we expect a form of task-performance as feedback from the environment. 
This feedback is modeled as an exponential function with slope ($k = 0.1$), consisting of a factor ($\textit{task}_{success} \in \{-1, 1\}$) and the time ($t$) to solve the task.
The task-performance score ($\mathcal{TP}_{Score}$) is calculated separately for each dimension of understanding (enabledness and comprehension). 
Consequently, our system requires two distinct task-performance measurements for the model.
The reward function $\mathcal{R}$ is normalized by both subtasks and scales with the positive reward factor $\lambda = 1$.
\vspace{-5pt} 

\begin{displaymath}
\begin{array}{ll}
\mathcal{TP}_{score} &= \textit{task}_{success} \cdot e^{-kT} \\ \\
\mathcal{R} &= \lambda \cdot \left( \frac{\mathcal{TP}_{\textit{compreh.}} + \mathcal{TP}_{\textit{enabl.}}}{2} \right)
\end{array}
\label{eq:reward_function}
\end{displaymath}

\section{Integration in Real-World Application}

\subsection{System Architecture}

All these theoretical concepts are integrated into a computational framework called \textbf{SHIFT} (Scaffolding, Human, Interdisciplinary, Framework, Tasks).
SHIFT is an adaptive framework that scaffolds human attention and understanding by combining interdisciplinary strategies, enhancing robotic explanations, and optimizing task execution in \ac{hri}.
The model has been implemented as a Python project and is available via docker\footnote{https://hub.docker.com/r/unibimas/shift}.
A video demonstration of the model in usage is available online\footnote{https://youtu.be/nRXk\_Tr3OJ8}.
The interfaces and specified \ac{ros} topics for querying the model can be found in the overview.
SHIFT is being developed as a three component architecture.
(1) \textbf{Computational model}: The computational part of the model as the systems' backend with the monitoring components, the dynamic state-space reduction and the learning algorithm.
(2) \textbf{Visualization}: A visual representation of human monitoring and the \ac{rl} in a web-based \ac{gui} using the nicegui framework for python \cite{schindler2024nicegui}.
(3) \textbf{Interface}: A \ac{ros} node as a server endpoint for external communication through \ac{ros} messages.
The endpoint implements a subscriber for human task-performance and eye-gaze behavior.
For ease of integration, all \ac{ros} messages are taken from the standard library, but are encoded using a JSON format.
In addition, the appropriate scaffolding strategy for a specific task can be queried through the model via a \ac{ros} service.
The computational part of the model is divided into three areas (\autoref{fig:model_overview}).
First, monitoring involves tracking the \textbf{partner model} by representing triples of observations during interactions.
Second, the \textbf{decision-making} system dynamically reduces the observed monitoring parameters and aligns them with human cognitive states.
The Q-learning algorithm then evaluates predicted actions for these cognitive states and selects the final action based on prior learning and the scoring system.
Third, \textbf{scaffolding} describes how the robot uses specific actions to guide the human in solving tasks.
During interactions, the environment provides task-performance, which serves as rewards for the learning agent. 
Finally, the monitoring components receive updated observation states from the environment in real-time again.

\subsection{Robotics Integration}

The presented system can be connected to any robotic architecture via \ac{ros}.
One tool for the integration of external components into \ac{hri} scenarios is RISE \cite{gross_schuetze_rise}.
RISE is a configurable scenario management component that makes it possible to describe robot behavior at a high level using simple \ac{xml} configurations.
The robots NAO \cite{naogouaillier2009mechatronic}, Pepper \cite{tanaka2015pepper} and the anthropomorphic robot head Flobi \cite{lutkebohle2010bielefeld} are currently supported by the system.
RISE offers to integrate external \ac{ros} nodes called feature nodes.
Our service component provides the scaffolding strategy for generating robot behavior with RISE and is integrated as a standalone feature within this architecture.
This allows for the autonomous querying of our service component during human-robot communication to determine the most suitable scaffolding strategy for the robot’s next task presentation, based on the human's current cognitive state.

\section{Evaluation: Model Adaptation}

The implementation of this model is driven by the need to identify the most effective scaffolding strategies for different cognitive states to enhance user task performance. 
Currently, no explicit correlations are known. 
As a first step, we evaluate the scoring system configuration to assess the impact of potential inaccuracies or mismatches with the user. 
We conduct simulations to test the model’s functionality and evaluate its effectiveness in social contexts.

\subsection{Parameter Optimization}

First, we are searching for the appropriate learning rate ($\alpha$), the discount factor ($\gamma$) and the initial exploration rate ($\epsilon$) for our setting.
We simulate a user (\textbf{User A}) who behaves optimally according to our configuration, and systematically change the learning parameters of our model.
Due to the limitation of interactions with real humans, we initially set a limit of 100 interactions (episodes) per simulation.
We focus on the cumulative reward ($\mathcal{R}$) and the time it takes for the cumulative reward to cross the zero line ($Z$).
It describes the time until the algorithm has recouped its cost of learning \cite{poole2010artificial}.
We run each simulation in this article 500 times and average the results (mean ($m$), standard deviation ($sd$)) to reduce environmental noise when rewarding (\autoref{tab:parameter_search}).
The discount factor describes the rate of taking future rewards into account and allows controlling rewards for sub processes until reaching a goal.
In the current setting, there is no planning component involved in the learning process.
The humans are solving tasks and receiving immediate rewards only.
The scoring system is one key component of our model, and the learning should not replace our initialization of the Q-table values immediately after receiving a negative reward.
Also, the exploration rate is decayed dynamically during runtime \cite{tokic2010adaptive} because of a limited time of learning at the beginning, due to the limited number of interactions.
We conduct multiple simulations with varied parameters, averaging results to reduce environmental noise in rewards, to end up with the final parameters $\alpha = 0.25$, $\gamma = 0.00$ and $\epsilon = 0.75$ for a first evaluation of the model.

\begin{table}[!ht]
\renewcommand{\arraystretch}{1.3}
\caption{Parameterization Model's Learning based on Pre-configured Scoring System ($\mathcal{H}_\mathcal{S} = \text{True}$) or without ($\mathcal{H}_\mathcal{S} = \text{False}$).}
\centering
\begin{tabular}{rcrrrrrrr}
  \toprule
 & $\mathcal{H}_\mathcal{S}$ & $\alpha$ & $\gamma$ & $\epsilon$ & $Z_{m}$ & $Z_{sd}$ & $\mathcal{R}_{m}$ & $\mathcal{R}_{sd}$ \\ 
  \hline
  \rowcolor{lightgray}
  1 & F & 0.25 & 0.00 & 0.75 & 31.78 & 32.90 & 13.39 & \textbf{5.30} \\ 
  \rowcolor{lightgray}
  2 & T & 0.25 & 0.00 & 0.75 & 13.99 & 20.66 & 19.62 & \textbf{4.71} \\
  3 & F & 0.25 & 0.50 & 0.75 & 32.82 & 33.62 & 12.87 & 5.36 \\ 
  4 & T & 0.25 & 0.50 & 0.75 & 13.18 & 20.04 & 19.98 & 4.98 \\ 
  5 & F & 0.25 & 0.95 & 0.75 & 31.94 & 34.60 & 9.82 & 6.32 \\ 
  6 & T & 0.25 & 0.95 & 0.75 & 15.25 & 22.43 & 19.07 & 5.07 \\ 
  7 & F & 0.50 & 0.00 & 0.75 & 33.91 & 35.12 & 12.11 & 5.36 \\ 
  8 & T & 0.50 & 0.00 & 0.75 & 14.10 & 21.02 & 18.33 & 4.95 \\ 
  9 & F & 0.50 & 0.50 & 0.75 & 31.76 & 33.72 & 12.86 & 5.64 \\ 
  10 & T & 0.50 & 0.50 & 0.75 & 13.47 & 21.09 & 18.77 & 5.00 \\ 
  11 & F & 0.50 & 0.95 & 0.75 & 34.03 & 38.20 & 7.26 & 6.95 \\ 
  12 & T & 0.50 & 0.95 & 0.75 & 14.92 & 21.19 & 18.73 & 5.04 \\ 
   \bottomrule
\end{tabular}
\vspace{-16pt}
\label{tab:parameter_search}
\end{table}

\subsection{Testing different Types of Users}

\begin{figure*}[!t]
    \centering
    \includegraphics[width=\linewidth]{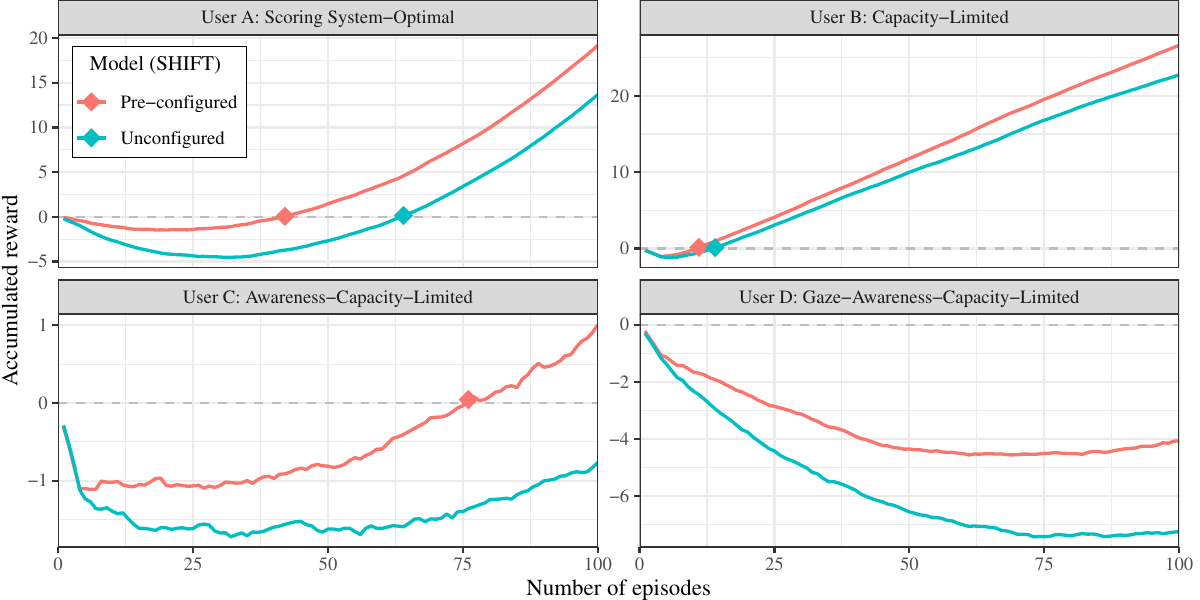}
    \vspace{-16pt} 
    \caption{Accumulated reward of Q-learning algorithm in 100 simulated interactions with and without the initialization of a Q-table based on the scoring system each. (A) is the optimal user, where  user performance corresponds to (B) 1-dimensional (processing capacity) (C) 2-dimensional (processing capacity and task awareness) (D) 3-dimensional (processing capacity, task awareness and gaze distribution) error in the pre-configuration for a negation strategy.}
    \label{fig:user_models}
    \vspace{-16pt} 
\end{figure*}

All simulations (\autoref{fig:user_models}) took place on the real system, and the individual subcomponents communicated in real time using \ac{ros} to challenge our system and test the pre-configuration of the system.
Therefore, we have modeled four types of users (A, B, C, D).
Each of them has a different behavior regarding our assumed pre-configuration.
\textbf{User A} faithfully follows our pre-configuration.
This user always performs well when the agent chooses an action that our pre-configuration would have suggested.
Users B, C, and D follow a different set of incorrect assumptions in the pre-configuration regarding the use of a negation strategy.
\textbf{User B} performs well when negation is used while processing capacity is low.
And performs bad when a negation is used when the capacity is high.
Our pre-configuration describes, for example, that a negation is preferred to use when the processing capacity is high (\autoref{tab:scoring_list}).
\textbf{User C} has the same attitude as user B regarding the processing capacity.
Furthermore, this user performs poorly when presented with a negation and has previously had a misconception on the enabledness side or has solved the task correctly.
\textbf{User D} follows the rules of user C.
In addition, even when presented with a negation, this user performs poorly when distracted or uncertain by gaze.
All user simulations include 5\% of behavior that deviates from the assigned user type.
In all four user simulations, the accumulated reward is higher with the use of the pre-configuration as initial Q-table.
User A tests whether the pre-configuration has any effect at all.
The pre-configured model adapts better overall than the model with a zero initialized values.
With the pre-configuration, the learning process recovers faster, requiring fewer episodes to regain efficiency ($\textit{episodes}_{pre} = 42 < \textit{episodes}_{un} = 63$).
The accumulated reward of user B crosses the zero line faster with the use of the pre-configuration ($\textit{episodes} = 11$) compared to the unconfigured model ($\text{episodes} = 14$).
In the simulation of user behavior C, the algorithm with the pre-configuration recouped at $\text{episodes} = 75$.
Without the pre-configuration, the algorithm fails to recover within the given timeframe.
For user D, all simulations do not recover within 100 episodes.

\subsection{Proof of Concept}

This evaluation serves as an initial proof of concept, assessing the model’s adaptability and learning capabilities in a social interaction context with less than 100 interactions. 
Since human-robot interactions are often brief, the ability to adapt within a limited number of exchanges is crucial for our model.
We introduced randomized user variations, allowing the model to test its scoring system-based decision-making in dynamic simulated interactions.
The results suggest that the pre-configured model improves adaptability, leading to faster recovery and better overall performance.
The scoring system is derived in a generalized manner from existing literature and prior interdisciplinary research.
We aimed to determine whether our model, initialized with this scoring system, could adapt when the scores were found to be inaccurate in certain dimensions.
Our findings underscore the potential of scoring system-guided adaptation in our context and serve as the foundation for future real-world experiments in \ac{hri}.

\section{Conclusion}

This work presents a method for formalizing a decision-making problem in robotics by integrating interdisciplinary research fields.
Therefore, we developed the SHIFT framework with its computational model for adaptively generating scaffolding strategies to guide humans in tasks in \ac{hri}. 
The model's adaptability was evaluated using a scoring system derived from interdisciplinary research, demonstrating its adaptability to different types of users and possible errors in the configuration.
We showed that the learning with our scoring system performs better in limited interactions from social contexts (higher acc. reward and faster recovery after exploration) than without it.
Future research will apply the model in \ac{hri} studies to improve the configuration of the scoring system and research on correlations between human cognitive states, represented by the partner model, and the use of verbal scaffolding strategies in explanatory dialogues.
We aim to explore how human input during runtime shape the configuration, potentially reducing the learning phase further and enhancing the applicability of our approach in social robotics.
Through this work, we lay the foundation for bridging interdisciplinary challenges with robotics, acknowledging the complexity in addressing these issues.
Despite limited research in this field, we strive to integrate diverse aspects into a comprehensive model, paving the way for future advancements.







\bibliographystyle{IEEEtran}
\bibliography{references}

\begin{thebibliography}{10}
\providecommand{\url}[1]{#1}
\csname url@samestyle\endcsname
\providecommand{\newblock}{\relax}
\providecommand{\bibinfo}[2]{#2}
\providecommand{\BIBentrySTDinterwordspacing}{\spaceskip=0pt\relax}
\providecommand{\BIBentryALTinterwordstretchfactor}{4}
\providecommand{\BIBentryALTinterwordspacing}{\spaceskip=\fontdimen2\font plus
\BIBentryALTinterwordstretchfactor\fontdimen3\font minus \fontdimen4\font\relax}
\providecommand{\BIBforeignlanguage}[2]{{%
\expandafter\ifx\csname l@#1\endcsname\relax
\typeout{** WARNING: IEEEtran.bst: No hyphenation pattern has been}%
\typeout{** loaded for the language `#1'. Using the pattern for}%
\typeout{** the default language instead.}%
\else
\language=\csname l@#1\endcsname
\fi
#2}}
\providecommand{\BIBdecl}{\relax}
\BIBdecl

\bibitem{sheridan2016human}
T.~B. Sheridan, ``Human--robot interaction: status and challenges,'' \emph{Human factors}, vol.~58, no.~4, pp. 525--532, 2016.

\bibitem{beetz2010cram}
M.~Beetz, L.~M{\"o}senlechner, and M.~Tenorth, ``Cram—a cognitive robot abstract machine for everyday manipulation in human environments,'' in \emph{2010 IEEE/RSJ international conference on intelligent robots and systems}.\hskip 1em plus 0.5em minus 0.4em\relax IEEE, 2010, pp. 1012--1017.

\bibitem{wang2018facilitating}
W.~Wang, R.~Li, Y.~Chen, Z.~M. Diekel, and Y.~Jia, ``Facilitating human--robot collaborative tasks by teaching-learning-collaboration from human demonstrations,'' \emph{IEEE Transactions on Automation Science and Engineering}, vol.~16, no.~2, pp. 640--653, 2018.

\bibitem{wang2018human}
W.~Wang, R.~Li, Y.~Chen, and Y.~Jia, ``Human intention prediction in human-robot collaborative tasks,'' in \emph{Companion of the 2018 ACM/IEEE international conference on human-robot interaction}, 2018, pp. 279--280.

\bibitem{tatarian2022does}
K.~Tatarian, R.~Stower, D.~Rudaz, M.~Chamoux, A.~Kappas, and M.~Chetouani, ``How does modality matter? investigating the synthesis and effects of multi-modal robot behavior on social intelligence,'' \emph{International Journal of Social Robotics}, vol.~14, no.~4, pp. 893--911, 2022.

\bibitem{Carlmeyer2014}
B.~Carlmeyer, D.~Schlangen, and B.~Wrede, ``Towards closed feedback loops in hri: Integrating inprotk and pamini,'' in \emph{Proceedings of the 2014 Workshop on Multimodal, Multi-Party, Real-World Human-Robot Interaction}, ser. MMRWHRI '14.\hskip 1em plus 0.5em minus 0.4em\relax ACM, 2014, pp. 1--6.

\bibitem{ren2015tfsm}
F.~Ren, Y.~Wang, and C.~Quan, ``Tfsm-based dialogue management model framework for affective dialogue systems,'' \emph{IEEJ Transactions on Electrical and Electronic Engineering}, vol.~10, no.~4, pp. 404--410, 2015.

\bibitem{zhao2019review}
Y.~J. Zhao, Y.~L. Li, and M.~Lin, ``A review of the research on dialogue management of task-oriented systems,'' in \emph{Journal of Physics: Conference Series}, vol. 1267.\hskip 1em plus 0.5em minus 0.4em\relax IOP Publishing, 2019, p. 012025.

\bibitem{KASNECI2023102274}
\BIBentryALTinterwordspacing
E.~Kasneci, K.~Sessler, S.~Küchemann, M.~Bannert, D.~Dementieva, F.~Fischer, U.~Gasser, G.~Groh, S.~Günnemann, E.~Hüllermeier, S.~Krusche, G.~Kutyniok, T.~Michaeli, C.~Nerdel, J.~Pfeffer, O.~Poquet, M.~Sailer, A.~Schmidt, T.~Seidel, M.~Stadler, J.~Weller, J.~Kuhn, and G.~Kasneci, ``Chatgpt for good? on opportunities and challenges of large language models for education,'' \emph{Learning and Individual Differences}, vol. 103, p. 102274, 2023. [Online]. Available: \url{https://www.sciencedirect.com/science/article/pii/S1041608023000195}
\BIBentrySTDinterwordspacing

\bibitem{rohlfing2020explanation}
K.~J. Rohlfing, P.~Cimiano, I.~Scharlau, T.~Matzner, H.~M. Buhl, H.~Buschmeier, E.~Esposito, A.~Grimminger, B.~Hammer, R.~H{\"a}b-Umbach \emph{et~al.}, ``Explanation as a social practice: Toward a conceptual framework for the social design of ai systems,'' \emph{IEEE Transactions on Cognitive and Developmental Systems}, vol.~13, no.~3, pp. 717--728, 2020.

\bibitem{kanda2017human}
T.~Kanda and H.~Ishiguro, \emph{Human-robot interaction in social robotics}.\hskip 1em plus 0.5em minus 0.4em\relax CRC Press, 2017.

\bibitem{almasri2019intelligent}
A.~Almasri, A.~Ahmed, N.~Almasri, Y.~S. Abu~Sultan, A.~Y. Mahmoud, I.~S. Zaqout, A.~N. Akkila, and S.~S. Abu-Naser, ``Intelligent tutoring systems survey for the period 2000-2018,'' \emph{IJARW}, 2019.

\bibitem{ramachandran2019personalized}
A.~Ramachandran, S.~S. Sebo, and B.~Scassellati, ``Personalized robot tutoring using the assistive tutor pomdp (at-pomdp),'' in \emph{Proceedings of the AAAI Conference on Artificial Intelligence}, vol.~33, no.~01, 2019, pp. 8050--8057.

\bibitem{naogouaillier2009mechatronic}
D.~Gouaillier, V.~Hugel, P.~Blazevic, C.~Kilner, J.~Monceaux, P.~Lafourcade, B.~Marnier, J.~Serre, and B.~Maisonnier, ``Mechatronic design of nao humanoid,'' in \emph{2009 IEEE international conference on robotics and automation}.\hskip 1em plus 0.5em minus 0.4em\relax IEEE, 2009, pp. 769--774.

\bibitem{martins2019alphapomdp}
G.~S. Martins, H.~Al~Tair, L.~Santos, and J.~Dias, ``$\alpha$pomdp: Pomdp-based user-adaptive decision-making for social robots,'' \emph{Pattern Recognition Letters}, vol. 118, pp. 94--103, 2019.

\bibitem{RAMIREZNORIEGA20171488}
\BIBentryALTinterwordspacing
A.~Ramírez-Noriega, R.~Juárez-Ramírez, and Y.~Martínez-Ramírez, ``Evaluation module based on bayesian networks to intelligent tutoring systems,'' \emph{International Journal of Information Management}, vol.~37, no. 1, Part A, pp. 1488--1498, 2017. [Online]. Available: \url{https://www.sciencedirect.com/science/article/pii/S0268401216302857}
\BIBentrySTDinterwordspacing

\bibitem{carlmeyer2018hesitating}
B.~Carlmeyer, S.~Betz, P.~Wagner, B.~Wrede, and D.~Schlangen, ``The hesitating robot-implementation and first impressions,'' in \emph{Companion of the 2018 ACM/IEEE International Conference on Human-Robot Interaction}, 2018, pp. 77--78.

\bibitem{folsom2013tractable}
J.~T. Folsom-Kovarik, G.~Sukthankar, and S.~Schatz, ``Tractable pomdp representations for intelligent tutoring systems,'' \emph{ACM Transactions on Intelligent Systems and Technology (TIST)}, vol.~4, no.~2, pp. 1--22, 2013.

\bibitem{kopp2018conversational}
S.~Kopp, M.~Brandt, H.~Buschmeier, K.~Cyra, F.~Freigang, N.~Kr{\"a}mer, F.~Kummert, C.~Opfermann, K.~Pitsch, L.~Schillingmann \emph{et~al.}, ``Conversational assistants for elderly users--the importance of socially cooperative dialogue,'' in \emph{Proceedings of the AAMAS Workshop on Intelligent Conversation Agents in Home and Geriatric Care Applications co-located with the Federated AI Meeting}, vol. 2338, 2018.

\bibitem{anwar2022applying}
A.~Anwar, I.~U. Haq, I.~A. Mian, F.~Shah, R.~Alroobaea, S.~Hussain, S.~S. Ullah, and F.~Umar, ``Applying real-time dynamic scaffolding techniques during tutoring sessions using intelligent tutoring systems,'' \emph{Mobile Information Systems}, vol. 2022, no.~1, p. 6006467, 2022.

\bibitem{wang2018reinforcement}
F.~Wang, ``Reinforcement learning in a pomdp based intelligent tutoring system for optimizing teaching strategies,'' \emph{International Journal of Information and Education Technology}, vol.~8, no.~8, pp. 553--558, 2018.

\bibitem{barraquand2008learning}
R.~Barraquand and J.~L. Crowley, ``Learning polite behavior with situation models,'' in \emph{Proceedings of the 3rd ACM/IEEE international conference on Human robot interaction}, 2008, pp. 209--216.

\bibitem{trr318_overall}
K.~J. Rohlfing and P.~Cimiano, ``Constructing explainability,'' in \emph{2022 IEEE 30th International Requirements Engineering Conference Workshops (REW)}, 2022, pp. 83--84.

\bibitem{booshehri2024towards}
M.~Booshehri, H.~Buschmeier, P.~Cimiano, S.~Kopp, J.~Kornowicz, O.~Lammert, M.~Matarese, D.~Mindlin, A.~S. Robrecht, A.-L. Vollmer \emph{et~al.}, ``Towards a computational architecture for co-constructive explainable systems,'' in \emph{Proceedings of the 2024 Workshop on Explainability Engineering}, 2024, pp. 20--25.

\bibitem{icaart23}
A.~Robrecht. and S.~Kopp., ``Snape: A sequential non-stationary decision process model for adaptive explanation generation,'' in \emph{Proceedings of the 15th International Conference on Agents and Artificial Intelligence - Volume 1: ICAART}, INSTICC.\hskip 1em plus 0.5em minus 0.4em\relax SciTePress, 2023, pp. 48--58.

\bibitem{gros_scaffolding_nodate}
A.~Groß, A.~Singh, N.~C. Banh, B.~Richter, I.~Scharlau, K.~J. Rohlfing, and B.~Wrede, ``Scaffolding the human partner by contrastive guidance in an explanatory human-robot dialogue,'' \emph{Frontiers in Robotics and AI}, vol.~10, 2023.

\bibitem{richter2023eeg}
B.~Richter, F.~Putze, G.~Ivucic, M.~Brandt, C.~Sch{\"u}tze, R.~Reisenhofer, B.~Wrede, and T.~Schultz, ``Eeg correlates of distractions and hesitations in human--robot interaction: a lablinking pilot study,'' \emph{Multimodal Technologies and Interaction}, vol.~7, no.~4, p.~37, 2023.

\bibitem{richter2021attention}
B.~Richter, ``The attention-hesitation model. a non-intrusive intervention strategy for incremental smart home dialogue management,'' Ph.D. dissertation, Bielefeld University, 2021.

\bibitem{klein2009erklaren}
J.~Klein, ``{Erklären-Was, Erklären-Wie, Erklären-Warum: Typologie und Komplexität zentraler Akte der Welterschließung},'' in \emph{{E}rkl{\"a}ren: {G}espr{\"a}chsanalytische und fachdidaktische {P}erspektiven}, R.~Vogt, Ed.\hskip 1em plus 0.5em minus 0.4em\relax Tübingen: Stauffenburg, 2009, pp. 25--36.

\bibitem{puterman1990markov}
M.~L. Puterman, ``Markov decision processes,'' \emph{Handbooks in operations research and management science}, vol.~2, pp. 331--434, 1990.

\bibitem{bundesen1990theory}
C.~Bundesen, ``A theory of visual attention.'' \emph{Psychological review}, vol.~97, no.~4, p. 523, 1990.

\bibitem{scharlau2020effects}
I.~Scharlau, A.~Kr{\"u}ger, K.~J. Rohlfing, and B.~Wrede, ``Effects of negation on visual processing capacity,'' in \emph{XPrag. de Workshop on the Processing of Negation and Polarity}, 2020.

\bibitem{Banh_Tünnermann_Rohlfing_Scharlau}
N.~C. Banh, J.~Tünnermann, K.~J. Rohlfing, and I.~Scharlau, ``Benefiting from binary negations? verbal negations decrease visual attention and balance its distribution,'' \emph{Frontiers in Psychology}, 2024, under review.

\bibitem{buschmeier2023formsunderstandingxaiexplanations}
\BIBentryALTinterwordspacing
H.~Buschmeier, H.~M. Buhl, F.~Kern, A.~Grimminger, H.~Beierling, J.~Fisher, A.~Groß, I.~Horwath, N.~Klowait, S.~Lazarov, M.~Lenke, V.~Lohmer, K.~Rohlfing, I.~Scharlau, A.~Singh, L.~Terfloth, A.-L. Vollmer, Y.~Wang, A.~Wilmes, and B.~Wrede, ``Forms of understanding of xai-explanations,'' 2023. [Online]. Available: \url{https://arxiv.org/abs/2311.08760}
\BIBentrySTDinterwordspacing

\bibitem{beltran2021inhibitory}
D.~Beltr{\'a}n, B.~Liu, and M.~de~Vega, ``Inhibitory mechanisms in the processing of negations: A neural reuse hypothesis,'' \emph{Journal of Psycholinguistic Research}, vol.~50, no.~6, pp. 1243--1260, 2021.

\bibitem{singh2023contrastiveness}
A.~Singh and K.~Rohlfing, ``Contrastiveness in the context of action demonstration: an eye-tracking study on its effects on action perception and action recall,'' in \emph{Proceedings of the Annual Meeting of the Cognitive Science Society}, vol.~45, no.~45, 2023.

\bibitem{collard2009disfluency}
P.~Collard, ``Disfluency and listeners' attention: An investigation of the immediate and lasting e ects of hesitations in speech,'' \emph{The University of Edinburgh}, 2009.

\bibitem{akalin2021reinforcement}
N.~Akalin and A.~Loutfi, ``Reinforcement learning approaches in social robotics,'' \emph{Sensors}, vol.~21, no.~4, p. 1292, 2021.

\bibitem{watkins1992q}
C.~J. Watkins and P.~Dayan, ``Q-learning,'' \emph{Machine learning}, vol.~8, pp. 279--292, 1992.

\bibitem{tokic2010adaptive}
M.~Tokic, ``Adaptive $\varepsilon$-greedy exploration in reinforcement learning based on value differences,'' in \emph{Annual conference on artificial intelligence}.\hskip 1em plus 0.5em minus 0.4em\relax Springer, 2010, pp. 203--210.

\bibitem{castro2014learning}
{\'A}.~Castro-Gonz{\'a}lez, M.~Malfaz, J.~F. Gorostiza, and M.~A. Salichs, ``Learning behaviors by an autonomous social robot with motivations,'' \emph{Cybernetics and Systems}, vol.~45, no.~7, pp. 568--598, 2014.

\bibitem{schindler2024nicegui}
\BIBentryALTinterwordspacing
F.~Schindler and R.~Trappe, ``Nicegui: Web-based user interfaces with python. the nice way.'' \url{https://github.com/zauberzeug/nicegui}, Jul. 2024, accessed: 2024-08-15. [Online]. Available: \url{https://github.com/zauberzeug/nicegui}
\BIBentrySTDinterwordspacing

\bibitem{gross_schuetze_rise}
A.~Gro{\ss}, C.~Sch{\"u}tze, M.~Brandt, B.~Wrede, and B.~Richter, ``Rise: An open-source architecture for interdisciplinary and reproducible hri research,'' \emph{Frontiers in Robotics and AI}, vol.~10, 2023.

\bibitem{tanaka2015pepper}
F.~Tanaka, K.~Isshiki, F.~Takahashi, M.~Uekusa, R.~Sei, and K.~Hayashi, ``Pepper learns together with children: Development of an educational application,'' in \emph{2015 IEEE-RAS 15th International Conference on Humanoid Robots (Humanoids)}.\hskip 1em plus 0.5em minus 0.4em\relax IEEE, 2015, pp. 270--275.

\bibitem{lutkebohle2010bielefeld}
\BIBentryALTinterwordspacing
I.~L{\"u}tkebohle, F.~Hegel, S.~Schulz, M.~Hackel, B.~Wrede, S.~Wachsmuth, and G.~Sagerer, ``The bielefeld anthropomorphic robot head “flobi”,'' in \emph{2010 IEEE International Conference on Robotics and Automation}, IEEE.\hskip 1em plus 0.5em minus 0.4em\relax Anchorage, Alaska: IEEE, 2010, pp. 3384--3391. [Online]. Available: \url{https://ieeexplore.ieee.org/document/5509173}
\BIBentrySTDinterwordspacing

\bibitem{poole2010artificial}
D.~L. Poole and A.~K. Mackworth, \emph{Artificial Intelligence: foundations of computational agents}.\hskip 1em plus 0.5em minus 0.4em\relax Cambridge University Press, 2010.

\end{thebibliography}

\end{document}